# Wave interactions and the analysis of the perturbed Burgers equation

Alex Veksler[1] and Yair Zarmi[1,2]
Ben-Gurion University of the Negev, Israel
[1]Department of Physics, Beer-Sheva, 84105
[2]Department of Energy & Environmental Physics
Jacob Blaustein Institute for Desert Research, Sede-Boqer Campus, 84990

Abstract

In multiple-front solutions of the Burgers equation, all the fronts, except for two, are generated through the inelastic interaction of exponential wave solutions of the Lax pair associated with the equation. The inelastically generated fronts are the source of two difficulties encountered in the standard Normal Form expansion of the approximate solution of the perturbed Burgers equation, when the zero-order term is a multiple-front solution: (*i*) The higher-order terms in the expansion are not bounded; (*ii*) The Normal Form (equation obeyed by the zero-order approximation) is not asymptotically integrable; its solutions lose the simple wave structure of the solutions of the unperturbed equation. The freedom inherent in the Normal Form method allows a simple modification of the expansion procedure, making it possible to overcome both problems in more than one way. The loss of asymptotic integrability is shifted from the Normal Form to the higher-order terms (part of which has to be computed numerically) in the expansion of the solution. The front-velocity update is different from the one obtained in the standard analysis.

## 1. Introduction

The Burgers equation [1] is the lowest order approximation for the one-dimensional propagation of weak shock waves in a fluid [2]. In recent years it has been employed in the description of the variation in vehicle density in highway traffic [3]. The equation is integrable. Its wave solutions are single-and multiple fronts. When a perturbation is added to the equation, its solutions are usually analyzed by the extension of the method of Normal Forms [4-6] to PDE's [7-12]. In the standard Normal Form analysis, the higher-order terms in the expansion are assumed to be differential polynomials of the zero-order solution, that is, they depend on the independent variables, $x$ and $t$, only through their dependence on the zero-order approximation.

When the zero-order approximation is a single-front, the approximate solution generated by the standard Normal Form analysis is an asymptotically integrable series with bounded higher-order terms. However, when the zero-order is not a single front, in the standard analysis, terms appear in the higher orders of the perturbative expansion that may not be accounted for by any differential polynomials in the asymptotic series [9, 10]. The only way to handle these terms in the standard analysis is to include them in the Normal Form (equation obeyed by the zero-order approximation) [7-12]. Unfortunately, this procedure yields a Normal Form that is not asymptotically integrable. Consequently, its solutions lose the simple wave structure of the solutions of the unperturbed equation; hence the name "obstacles to asymptotic integrability". Obstacles to integrability have been discussed extensively in the literature, in the cases of the perturbed KdV and NLS equations [7, 8, 11, 12]. It has been observed that obstacles to integrability in the Normal Form analysis are consequences of inelastic wave interactions [12].

In addition to the obstacles to integrability, the standard Normal Form analysis of the perturbed Burgers equation suffers from another serious problem: Some of the contributions to the higher-order terms in the approximate solution that are allowed by the formalism and can be computed in closed form, become unbounded in the multiple-wave case. The two difficulties enumerated above limit the capacity of the standard analysis to generate asymptotically integrable perturbative expansions with bounded higher-order terms to a subset of the possible perturbations.

The source of both problems will be shown to be the inelastic interaction of exponential wave solutions of the Lax pair [13-14] associated with the Burgers equation. First, unbounded terms occurring in higher orders in the expansion of the standard analysis emerge just along the inelastically generated fronts. Second, the mere existence of inelastic interactions leads to the emergence of obstacles to integrability. The freedom available in the standard analysis enables one to confine the effect of the obstacles to these fronts, along which, the obstacles asymptote to symmetries of the Burgers equation. As a result, even if a way is found that allows one to handle the obstacles to integrability within the series expansion of the solution rather than in the Normal Form, they generate secular behavior in the solution.

The undesired effects of inelastic wave interactions, encountered in the standard analysis, can be avoided by exploiting the freedom in the expansion. The assumption, usually made in the standard analysis [7-12], that the higher-order corrections to the approximate solution of the perturbed equation depend on the independent variables, $x$ and $t$ only through their dependence on the zero-order approximation, is not a pre-requisite of the Normal Form expansion procedure. Allowing higher-order corrections to depend explicitly on $x$ and $t$, additional freedom is gained, which enables one to eliminate the undesired effects in more than one way.

Whenever a secular term is generated in the expansion, this is a sign that a "hidden" symmetry (i.e., a term that asymptotes to a symmetry) has been missed. We identify these hidden symmetries, and assign them to the Normal Form. This takes care of unbounded terms generated either directly in the expansion, or through the effect of obstacles. This procedure yields an update of front-velocities that is different from the one obtained through the standard analysis.

The added freedom allows one to shift the terms that spoil integrability (e.g., remnants of obstacles, with the "hidden" symmetry contribution removed from them) from the Normal Form to the higher-order terms in the expansion of the solution. The loss of asymptotic integrability is inevitable, requiring that parts of the solution have to be computed numerically. However, the Normal Form remains asymptotically integrable. Hence, its solutions retain the simple wave structure of the solutions of the unperturbed equation.

In Section 2, we review the properties of the wave-front solutions of the unperturbed Burgers equation. In the multiple-front case, the fronts are constructed from exponential wave solutions of the Lax pair associated with the Burgers equation. Two of the fronts are generated each from one exponential wave solution of the Lax pair, and are unaffected by the existence of other waves. All the remaining fronts are generated through the inelastic interaction of pairs of exponential waves with adjacent wave numbers.

In Section 3, the known properties of the symmetries of the Burgers equation are reviewed. In Section 4, we review the Normal Form analysis of the perturbed equation. We show how obstacles to asymptotic integrability and unbounded (secular) terms emerge in the multiple-wave case, just along the inelastically generated fronts. In Section 5, we present a number of aspects of the freedom inherent in the perturbative expansion, including the freedom introduced due to the addition of explicit $x$ and $t$ dependence in the higher-order terms in the perturbation series.

The remainder of the paper is dedicated to the exploitation of that freedom for circumventing the difficulties encountered in the standard analysis. We show two ways to achieve this goal through first order in the perturbation. In Section 6, the freedom is exploited to rigorously eliminate both obstacles to integrability as well as secular terms in the perturbation series. The price paid is that some contributions in the first-order correction to the solution have to be computed numerically. We find that these terms are bounded in both the single- and two-wave cases. They are non-negligible and persist asymptotically away from the origin. The approximate solutions of the standard analysis and the alternative suggested in Section 6 are compared in Section 7. Within the error allowed by the perturbation formalism, they are equivalent.

In Section 8, we present another alternative for circumventing obstacles to integrability and secular terms in the multiple-wave case. We exploit the fact that, away from the origin, all the fronts in a multiple-wave solution asymptote to well separated single fronts at an exponentially fast rate and that, for single fronts, the expansion generates an integrable series with bounded terms. The freedom in the expansion allows us to construct an asymptotically integrable Normal Form separately for each front. The numerical corrections in the first-order term in the expansion become exponentially small as the distance from the origin grows.

**2. Inelastic wave interactions in the Burgers equation**

The wave solutions of the Burgers equation [1]

$$u_t = 2u u_x + u_{xx} \qquad (2.1)$$

are fronts. The commonly studied solution is that of a front with one vanishing boundary value:

$$u(t,x) = \frac{k e^{k(x+kt+x_0)}}{1 + e^{k(x+kt+x_0)}} \qquad (2.2)$$

The general single-front solution, with nonvanishing boundary values at both $x = \pm 8$, will surface in the analysis of multiple-wave solutions. It is obtained from Eq. (2.2) by a Galilean transformation, and has the form

$$u(t,x) = \frac{u_m + u_p e^{k(x+vt+x_0)}}{1 + e^{k(x+vt+x_0)}} \quad , \quad v = u_p + u_m \, , \, k = u_p - u_m \qquad (2.3)$$

The fronts, Eqs. (2.2) and (2.3), are generated by exponential wave solutions of the linearization of Eq. (2.1) through the Lax pair [13,14] associated with the Burgers equation, given by [15-18]. For the solution of Eq. (2.2), the Lax pair is

$$y_x = u\,y \tag{2.4}$$

$$y_t = y_{xx} \tag{2.5}$$

Linear combinations of wave solutions of this Lax pair generate multiple-wave solutions of the Burgers equation, given by

$$u(t,x) = \frac{\sum_{i=1}^{M} k_i\, e^{k_i (x + k_i t + x_{i,0})}}{1 + \sum_{i=1}^{M} e^{k_i (x + k_i t + x_{i,0})}} \tag{2.6}$$

For later use, we define the characteristic line of each exponential wave solution by

$$x = -k_i\, t - x_{i,0} \tag{2.7}$$

$u(t,x)$ of Eq. (2.6) vanishes asymptotically in a triangular wedge in the $x - t$ plane, defined by some pair of two adjacent characteristic lines. The direction in which the solution vanishes depends on the values of the wave numbers, $k_i$. Without loss of generality, we assume that all wave numbers are positive, and arrange them in ascending order:

$$0 < k_1 < k_2 < \ldots < k_M \tag{2.8}$$

The solution then vanishes between the characteristic lines defined by $k_1$ and $k_M$. (If some of the wave numbers are negative, then the zero-boundary value is common to the fronts generated by $k_-$, the wave number with highest negative value, by $k_+$, the wave number with lowest positive value.)

As a solution with non-zero boundary values can always be transformed into one with one vanishing boundary value by a Galilean transformation, we focus on the solution given in Eq. (2.6).

The single-wave solution describes a front of infinite extent along its characteristic line. The solution with $M > 1$ waves (Eq. (2.6)) describes $(M + 1)$ semi-infinite fronts, each extending along one half of some characteristic line. All the fronts collide in the vicinity of the origin. The locations of the collision points depend on the shifts $x_{i,0}$. The collision region is a finite domain around the origin in the $x - t$ plane. Fig. 1 shows examples of single- and three-wave solution.

The multiple-wave solution has remarkable properties: (*i*) Away from the collision region all fronts asymptote exponentially fast to *single-front solutions*; (*ii*) Two of the asymptotic fronts have one vanishing boundary condition. They are not affected by the existence of other waves. Each of the two fronts is generated from one of the exponential wave solutions appearing in Eq. (2.6). Their characteristic lines are given by Eq. (2.7). (*iii*) The remaining $(M - 1)$ fronts are generated through the inelastic interaction of exponential waves with *adjacent* wave numbers. Each front has two nonzero boundary conditions.

As an example, we analyze the two-wave solution, given by

$$u(t,x) = \frac{k_1 e^{k_1(x + k_1 t + x_{1,0})} + k_2 e^{k_2(x + k_2 t + x_{2,0})}}{1 + e^{k_1(x + k_1 t + x_{1,0})} + e^{k_2(x + k_2 t + x_{2,0})}} \tag{2.9}$$

Eq. (2.9) generates three semi-infinite fronts. Two of the fronts evolve along the characteristic lines of the two exponential waves. Away from the origin, they asymptote to single-front solutions, each with one vanishing boundary value, of the form of Eq. (2.2):

$$u(t,x) \xrightarrow[t \to +\infty, \ x \cong -k_2 t - x_{2,0}]{} \frac{k_2 e^{k_2(x + k_2 t + x_{2,0})}}{1 + e^{k_2(x + k_2 t + x_{2,0})}} \tag{2.10a}$$

$$u(t,x) \xrightarrow[t \to -\infty,\ x \cong -k_1 t - x_{1,0}]{} \frac{k_1 e^{k_1(x + k_1 t + x_{1,0})}}{1 + e^{k_1(x + k_1 t + x_{1,0})}} \qquad (2.10b)$$

The third semi-infinite front is generated by the inelastic interaction of the two exponential waves in Eq. (2.9), with a characteristic line defined by the equality of the two exponents appearing in Eq. (2.9):

$$k_1(x + k_1 t + x_{1,0}) = k_2(x + k_2 t + x_{2,0}) \Rightarrow x = -(k_1 + k_2)t - \mathbf{x} \qquad \left(\mathbf{x} = \frac{k_2 x_{2,0} - k_1 x_{1,0}}{k_2 - k_1}\right) \qquad (2.11)$$

It asymptotes to a single-front solution, of the type given by Eq. (2.3),

$$u(t,x) \xrightarrow[t \to -\infty,\ x \cong -(k_1 + k_2)t - \mathbf{x}]{} \frac{k_1 + k_2 e^{(k_2 - k_1)(x + (k_1 + k_2)t + \mathbf{x})}}{1 + e^{(k_2 - k_1)(x + (k_1 + k_2)t + \mathbf{x})}} \qquad (2.12)$$

The boundary values of the asymptotic front described by Eq. (2.12) are both non-zero: $k_1$ and $k_2$ for $x \to -\infty$ and $+\infty$, respectively, its wave number is $k_2 - k_1$, and its velocity is $k_2 + k_1$.

The picture in the case of $M$ exponential waves is similar. The waves with wave numbers $k_1$ and $k_M$ generate semi-infinite fronts that asymptote to single fronts with one vanishing boundary value, given by Eq. (2.2). One front has $k = k_1$, and the other has $k = k_M$. The remaining $(M - 1)$ semi-infinite fronts are generated by the inelastic interaction of waves with adjacent wave numbers, $k_i$ and $k_{i+1}$ ($1 < i < M - 1$). They asymptote to single fronts, with both boundary values non-vanishing, $k_i$ and $k_{i+1}$, wave numbers $(k_{i+1} - k_i)$ and velocities $(k_{i+1} + k_i)$.

## 3. Symmetries of the Burgers equation

As we shall use the symmetries of the Burgers equation repeatedly, we summarize their properties [16-18] in brief. The symmetries, $S_n$, are solutions of the linearization of the Burgers equation

$$\partial_t S_n = 2\partial_x(u S_n) + \partial_x^2 S_n \qquad (3.1)$$

Here $u$ is a solution of Eq. (2.1). The hierarchy of symmetries obeys the recursion relation

$$S_{n+1} = \partial_x \{S_n + uG_n\} \tag{3.2}$$

with

$$G_n[u] = \partial_x^{-1} S_n[u], \qquad G_0 = 1 \tag{3.3}$$

Eqs. (3.2) and (3.3) yield a relation that will turn out to be useful in the following:

$$S_n = G_{n+1} - G_1 G_n \tag{3.4}$$

For the present analysis, we shall need the first three symmetries

$$\begin{aligned} G_1 &= u, & S_1 &= u_x \\ G_2 &= u^2 + u_x, & S_2 &= 2u u_x + u_{xx} \\ G_3 &= u^3 + 3u u_x + u_{xx}, & S_3 &= 3u^2 u_x + 3u u_{xx} + 3u_x^2 + u_{xxx} \end{aligned} \tag{3.5}$$

The Lie brackets of any two symmetries vanish

$$[S_m[u], S_n[u]] \equiv \sum_i \left\{ \frac{\partial S_m}{\partial u_i} \partial_x^i S_n - \frac{\partial S_n}{\partial u_i} \partial_x^i S_m \right\} = 0 \qquad (u_i \equiv \partial_x^i u). \tag{3.6}$$

When $u$ is the single-front solution, Eq. (2.3), the hierarchy degenerates into a single symmetry. Using the recursion relation, Eq. (3.2) one can prove by induction that

$$S_n[u] = c_n S_1[u] \qquad \left\{ c_n = \frac{(u_p)^n - (u_m)^n}{u_p - u_m} \right\} \tag{3.7}$$

$$G_n[u] = c_n G_1[u] + u_m^n - c_n u_m \tag{3.8}$$

If one of the boundary values vanishes, then one also has

$$G_n[u] = c_n G_1[u] \quad , \quad c_n = k^{n-1} \tag{3.9}$$

In the discussion of obstacles to asymptotic integrability, we shall encounter the entity

$$R_{21} = S_2[u]G_1[u] - S_1[u]G_2[u] \tag{3.10}$$

A trivial consequence of Eqs. (3.7) and (3.9) is that $R_{21}$ vanishes identically when $u$ is a single-front solution with one vanishing boundary value.

## 4. Difficulties in standard Normal Form expansion of perturbed Burgers equation

Through first order, the perturbed Burgers equation is given by

$$\begin{aligned} w_t &= 2ww_x + w_{xx} \\ &+ \boldsymbol{e}\left(3\boldsymbol{a}_1 w^2 w_x + 3\boldsymbol{a}_2 ww_{xx} + 3\boldsymbol{a}_3 w_x^2 + \boldsymbol{a}_4 w_{xxx}\right) \end{aligned} \tag{4.1}$$

One now expands the solution in a Near Identity Transformation (NIT)

$$w = u + \boldsymbol{e} u^{(1)} + O(\boldsymbol{e}^2) \tag{4.2}$$

One then allows for the possibility of including part of the effect of the perturbation in the evolution of the zero-order approximation, $u$, by imposing that it obeys a Normal Form (NF), given by

$$u_t = 2uu_x + u_{xx} + \boldsymbol{e} U_1 + O(\boldsymbol{e}^2) \tag{4.3}$$

Inserting Eqs. (4.2) and (4.3) in Eq. (4.1), 0ne obtains the first-order *homological equation*:

$$\begin{aligned} U_1 + \partial_t u^{(1)} &= 2\partial_x\left(uu^{(1)}\right) + \partial_x^2 u^{(1)} + \\ &\quad \left(3\boldsymbol{a}_1 u^2 u_x + 3\boldsymbol{a}_2 uu_{xx} + 3\boldsymbol{a}_3 u_x^2 + \boldsymbol{a}_4 u_{xxx}\right) \end{aligned} \tag{4.4}$$

To generate both $u^{(1)}$ and $U_1$ from Eq. (4.4), one adds the requirement that $u^{(1)}$ must not develop unbounded (secular) behavior. This is obtained by assigning the resonant part in the driving term in Eq. (4.4) to $U_1$, so that it is moved from Eq. (4.4) to the NF, Eq. (4.3). The resonant term is proportional to the symmetry $S_3$:

$$U_1 = mS_3[u] = m\left(3u^2 u_x + 3u u_{xx} + 3u_x^2 + u_{xxx}\right) \qquad (4.5)$$

The value of the coefficient **m** depends on the specific case.

With this form of $U_1$, Eq. (4.3) is integrable. Its wave solutions are the single- and multiple-front solutions of the unperturbed equation, Eq. (2.1), with one modification: the velocities are updated due to the effect of the perturbation. In addition, properties of the symmetries, derived for the case when $u$ was the solution of Eq. (2.1), hold also when $u$ is a solution of the Normal Form, provided $U_1$ is a symmetry. This can be shown to be a consequence of the fact that the symmetries, $S_n$, are differential polynomials in $u$ and its spatial derivatives. In particular, Eqs. (3.7)-(3.9) hold when $u$ is a single-wave solution.

With $U_1$ of Eq. (4.5), Eq. (4.4) becomes

$$\partial_t u^{(1)} = 2\partial_x\left(u u^{(1)}\right) + \partial_x^2 u^{(1)} + \\ \left(3(a_1 - m)u^2 u_x + 3(a_2 - m)u u_{xx} + 3(a_3 - m)u_x^2 + (a_4 - m)u_{xxx}\right) \qquad (4.6)$$

In the standard analysis, $u^{(1)}$ is assumed to be a functional of the zero-order term, $u$. The formalism then allows for $u^{(1)}$ the following structure [9,10]:

$$u^{(1)}[u] = au^2 + bqu_x + cu_x \quad \left(q \equiv \partial_x^{-1} u\right) \qquad (4.7)$$

The coefficient $c$ is undetermined, as it multiplies a symmetry, the contribution of which in the homological equation vanishes.

The ability to solve Eq. (4.6) for $u^{(1)}$ crucially depends on the structure of the zero-order term, $u$. Consider, first, the single-front solution of Eq. (2.3). The coefficients $a$ and $b$ can be solved for, and have the values [9,10]

$$a = -\tfrac{1}{2}(2\mathbf{a}_1 - \mathbf{a}_2 + \mathbf{a}_3 - 2\mathbf{a}_4) \quad , \quad b = (\mathbf{a}_1 - 2\mathbf{a}_2 - \mathbf{a}_3 + 2\mathbf{a}_4) \tag{4.8}$$

The solution of the normal form remains a single-wave one. The coefficient $m$ obtains the value

$$m = \frac{(u_m u_p)(\mathbf{a}_1 + \mathbf{a}_2 - \mathbf{a}_3) + (u_m^2 + u_p^2)\mathbf{a}_4}{u_m^2 + u_m u_p + u_p^2} \tag{4.9}$$

As a result, the velocity is updated by the perturbation according to

$$v = u_m + u_p + e\left\{(u_m u_p)(\mathbf{a}_1 + \mathbf{a}_2 - \mathbf{a}_3) + (u_m^2 + u_p^2)\mathbf{a}_4\right\} \tag{4.10}$$

If one of the boundary values vanishes, a simpler update is obtained:

$$m = \mathbf{a}_4 \tag{4.11}$$

$$v = k + \mathbf{a}_4 k^2 \tag{4.12}$$

Note that $q = \partial_x^{-1} u$, diverges linearly:

$$q[t,x] = \frac{u_p}{k}\ln\left\{1 + e^{k(x+vt+x_0)}\right\} - \frac{u_m}{k}\ln\left\{1 + e^{-k(x+vt+x_0)}\right\} \tag{4.13}$$

However, the term $q \cdot u_x$ in Eq. (4.7) is bounded, because $u_x$ falls off exponentially fast away from the characteristic line of the front.

We now turn to the general case, when the zero-order term, $u$, is not a single-wave solution. The standard Normal Form analysis then encounters an obstacle to integrability. With $U_1$ given by Eq. (4.5), we focus on the multiple-wave solutions of Eq. (4.3) a modified version of Eq. (2.6):

$$u(t,x) = \frac{\sum_{i=1}^{M} k_i e^{k_i(x + v_i t + x_{i,0})}}{1 + \sum_{i=1}^{M} e^{k_i(x + v_i t + x_{i,0})}} \qquad v_i = v_{i,0} + \mathbf{e}\,\mathbf{m} k_i^2 \tag{4.14}$$

As the solution given in Eq. (4.14) has one vanishing boundary condition, the choice of Eq. (4.11) is adopted. Eq. (4.6) then becomes

$$\partial_t u^{(1)} = 2\partial_x\!\left(u u^{(1)}\right) + \partial_x^2 u^{(1)} + \\ \left(3(\mathbf{a}_1 - \mathbf{a}_4) u^2 u_x + 3(\mathbf{a}_2 - \mathbf{a}_4) u u_{xx} + 3(\mathbf{a}_3 - \mathbf{a}_4) u_x^2\right) \tag{4.15}$$

Inserting Eqs. (4.7) and (4.11) in Eq. (4.15), $a$ and $b$ have to account for three independent contributions in the driving term in Eq. (4.15). As a result, $a$ and $b$ cannot be determined unless [9,10]

$$\mathbf{g} \equiv 2\mathbf{a}_1 - \mathbf{a}_2 - 2\mathbf{a}_3 + \mathbf{a}_4 = 0 \tag{4.16}$$

As there is no way to determine the coefficients $a$ and $b$, one has to make a choice. We choose for them the values found in the single-front case, given in Eq. (4.8). With that choice, the terms that remain unaccounted for are the ones that would automatically vanish in the single-wave case. With $R_{21}$ defined in Eq. (3.9), Eq. (4.15) is reduced to

$$0 = \mathbf{g}\left(u^2 u_x + u u_{xx} - u_x^2\right) = \mathbf{g}\,R_{21}[u] \tag{4.17}$$

If $\mathbf{g} \ne 0$, there is no way to account for $R_{21}$ by $u^{(1)}$. One is forced to shift it into $U_1$:

$$U_1 = \mathbf{a}_4 S_3[u] + \mathbf{g}\,R_{21} \tag{4.18}$$

As a result, the Normal Form, Eq. (4.3) becomes

$$u_t = S_2[u] + e(a_4 S_3[u] + R_{21}[u])_1 + O(e^2) \qquad (4.19)$$

As $R_{21}$ is not a symmetry, it spoils the integrability of the Normal Form, and the solutions lose the simple structure of those of the unperturbed equation. As a consequence, the properties of the symmetries, $S_n$ reviewed at the end of Section 3 may not hold.

Concurrently with the loss of asymptotic integrability, when $u$ is a multiple-wave solution, another serious problem is hidden in Eq. (4.7): The term $(b \cdot q \cdot u_x)$, which is bounded in the single-front case, develops a linearly divergent (secular) behavior along the characteristic lines of the inelastically generated fronts. In the case of a two-wave solution, Eq. (2.9), one has

$$q(t,x) = \ln\left\{ + e^{k_1(x + k_1 t + x_{1,0})} + e^{k_2(x + k_2 t + x_{2,0})} \right\} \qquad (4.20)$$

Along the characteristic line of the inelastically generated front (see Eq. (2.11)), $q \cdot u_x$, diverges linearly, making $u^{(1)}$ unbounded, because leading term in the asymptotic behavior of $q(t,x)$ is:

$$q_0(t) \equiv q(t,x)\big|_{x \to -(k_1 + k_2)t - x} \xrightarrow[k_1 k_2 t \to +\infty]{} -k_1 k_2 t \qquad (4.21)$$

($q_0$ vanishes exponentially fast for $k_1 \cdot k_2 \cdot t \to -\infty$). To ensure that the standard analysis generates a bounded first-order approximation, term $q \cdot u_x$ cannot be included in Eq. (4.7). This leads to the emergence of an additional contribution to the obstacle to integrability:

$$-(a_1 - 2a_2 - a_3 + 2a_4)\{2R_{21} - u^2 u_x + 2u_x^2\} \qquad (4.22)$$

Note that the numerical coefficient in Eq. (4.22) is the value of $b$ obtained in the single-front case (Eq. (4.8)). Thus, another option is to require that

$$b = a_1 - 2a_2 - a_3 + 2a_4 = 0 \tag{4.23}$$

If, in addition, one imposes the requirement that no obstacles to integrability emerge (Eq. (4.16)), the class of perturbations for which the standard method generates an asymptotically integrable expansion with bounded higher-order terms is limited even further, to:

$$a_1 = a_3 \quad , \quad a_2 = a_4 \tag{4.24}$$

## 5. Freedom in the Normal Form Expansion

The freedom inherent in the Normal Form expansion is not fully exploited in the standard analysis. In particular, the assumption that the higher-order corrections do not depend explicitly on the independent variables, $x$ and $t$, but only through their dependence the zero-order solution, $u$, is not a fundamental requirement of the method. This assumption eventually forces one to include the obstacle to integrability in the Normal Form, (see Eq. (4.19)). Thus, let us allow the first-order correction, $u^{(1)}$, to contain an explicitly $t$- and $x$-dependent term:

$$u^{(1)}[u;t,x] = au^2 + bqu_x + cu_x + w^{(1)}(t,x) \tag{5.1}$$

The terms that are unaccounted for by the differential polynomial part of $u^{(1)}$ can be accounted for by $w^{(1)}(t,x)$, preventing the need to transfer the obstacle to integrability to the Normal Form.[1]

With the inclusion of $w^{(1)}(t,x)$ in Eq. (5.1), one has additional freedom in dividing the effect of the perturbation between the Normal Form and the Near Identity Transformation, i.e., between $U_1$ and $u^{(1)}$. One clearly wishes to identify terms that, if retained in the homological equation, may gener-ate secular behavior in $u^{(1)}$. To this end, it pays to re-write the homological equation, Eq. (4.4), in terms of independent quantities that are identified as having or not having the capacity to generate secular behavior, rather than in terms of the traditional monomials $u^2 u_x$, $u \cdot u_{xx}$, $u_x^2$ and $u_{xxx}$. The symmetry, $S_3$, generates secular behavior. It will be shown in Section 6, that the obstacle $R_{21}$, although not a symmetry, has the potential of generating a secular contribution in $u^{(1)}$ along inelastically generated fronts. The two differential polynomials, $?_x(u \cdot u_x)$ and $?_x S_2[u]$, will be shown to generate bounded behavior in $u^{(1)}$. Inserting Eq. (5.1) in Eq. (4.4), the latter can be recast in the following form:

$$U_1 + \partial_t w^{(1)} = 2\partial_x \left(u w^{(1)}\right) + \partial_x^2 w^{(1)} + mS_3[u] + l\, R_{21}[u] + r\partial_x(u u_x) + s\, \partial_x S_2[u] \qquad (5.2)$$

with

$$m = a + a_1 - \tfrac{1}{2}a_2 + \tfrac{1}{2}a_3 \qquad (5.3)$$

$$l = -a + b + \tfrac{3}{2}a_2 - \tfrac{3}{2}a_3 \qquad (5.4)$$

$$r = b - a_1 + 2a_2 + a_3 - 2a_4 \qquad (5.5)$$

$$s = a_4 - m \qquad (5.6)$$

---

[1] The problem arises already in the Normal Form analysis of perturbed ODE's, where, customarily, higher-order corrections in the Near Identity Transformation (NIT) are assumed to have no explicit time dependence. This assumption works for systems of *autonomous* equations with a *linear unperturbed part*. Inconsistencies may emerge in all other cases, and are resolved by allowing for explicit time dependence in the NIT.

## 6. Elimination of obstacle to integrability and secular behavior

Of the four differential polynomials in Eq. (5.2), the symmetry, $S_3$, will generate a secular term in $w^{(1)}$, hence it must be accounted for by $U_1$

$$U_1 = mS_3[u] \qquad (6.1)$$

The remaining terms are not symmetries, hence it pays not to include them in $U_1$.

The obstacle, $R_{21}$, does not emerge when zero-order approximation, $u$, is a single-wave. However, it does emerge in the case of a general solution for $u$. Consider the case when $u$ is an $M$- wave solution, which exhibits $(M + 1)$ semi-infinite fronts. Away from the collision region, the semi-infinite fronts asymptote to single-fronts at an exponential rate (see Section 2). Of these fronts, two have one vanishing boundary value. As $R_{21}$ vanishes explicitly for a single-wave case with one vanishing boundary value (see Section 3), in the multiple-wave case, along these two fronts, $R_{21}$ is expected to decay to zero at the same exponential rate. The remaining $(M - 1)$ fronts all have two non-zero boundary values (see Section 2). As a result, $R_{21}$ does not vanish, along these fronts.

To be specific, let us focus on the case of the two-wave solution of Eq. (2.9), with the inelastically generated front defined by Eq. (2.11). $R_{21}$ asymptotes to

$$R_{21}[u] \to \begin{Bmatrix} 0 \\ k_1 k_2 u_x \end{Bmatrix} + e.s.t \qquad \begin{matrix} x \to -k_i t - x_{i,0}, \quad (i = 1, 2) \\ \\ x \to -(k_1 + k_2)t - \mathbf{x} \end{matrix} \qquad (6.2)$$

In Eq. (6.2), "e.s.t." stands for small terms that vanish at an exponential rate along the characteristic line of the front as the distance from the origin grows. In the $M$-wave case, $k_1$ and $k_2$ are replaced by $k_i$ and $k_{i+1}$, respectively, for each of the $(M - 1)$ inelastically generated fronts.

The obstacle is not a symmetry, i.e., it is not a solution of Eq. (3.1). However, it asymptotes to a symmetry along the inelastically generated fronts. Therefore, in Eq. (5.2), it is expected to generate secular terms in $w^{(1)}$ along these characteristic lines. One way to avoid this, is to choose

$$\mathbf{l} = 0 \qquad (6.3)$$

Towards the end of Section 4, it was pointed out that, in the multiple-wave case, the term $q \cdot u_x$ has a secular behavior. Thus, it must be eliminated from $u^{(1)}$. This requires

$$b = 0 \qquad (6.4)$$

Using Eqs. (6.3) and (6.4) in Eqs. (5.3)-(5.6), yields

$$a = \tfrac{3}{2}\mathbf{a}_2 - \tfrac{3}{2}\mathbf{a}_3 \tag{6.5}$$

$$\mathbf{m} = \mathbf{a}_1 + \mathbf{a}_2 - \mathbf{a}_3 \tag{6.6}$$

$$\mathbf{r} = -\mathbf{a}_1 + 2\mathbf{a}_2 + \mathbf{a}_3 - 2\mathbf{a}_4 \tag{6.7}$$

$$\mathbf{s} = -\mathbf{a}_1 - \mathbf{a}_2 + \mathbf{a}_3 + \mathbf{a}_4 \tag{6.8}$$

With these values for the coefficients, the Normal Form, Eq. (4.3), becomes

$$u_t = S_2[u] + e(\mathbf{a}_1 + \mathbf{a}_2 - \mathbf{a}_3)S_3[u] + O(e^2) \tag{6.9}$$

Eq. (6.9) is integrable. Its solutions are the single- and multiple-wave solutions of Eq. (2.1), except that, instead of Eq. (4.12), the velocity of each wave is now updated according to

$$v_i = k_i + e(\mathbf{a}_1 + \mathbf{a}_2 - \mathbf{a}_3)k_i^2 \tag{6.10}$$

In addition, Eq. (5.2) is reduced to

$$\partial_t w^{(1)} = 2\partial_x\left(u\,w^{(1)}\right) + \partial_x^2 w^{(1)} + \\ \left(-\mathbf{a}_1 + 2\mathbf{a}_2 + \mathbf{a}_3 - 2\mathbf{a}_4\right)\partial_x(uu_x) + \left(-\mathbf{a}_1 - \mathbf{a}_2 + \mathbf{a}_3 + \mathbf{a}_4\right)\partial_x S_2[u] \tag{6.11}$$

A proof that $?_x(u \cdot u_x)$ and $?_x S_2[u]$ do not generate secular behavior in $w^{(1)}$ in Eq. (6.11) is still missing. Numerically, we have found that they do not generate such behavior. We have checked this for single- and two-wave solutions for various boundary conditions and integration domains. Examples of the numerical solutions of Eq. (6.11) with either driving term are shown in Figs. 2 & 3

## 7. Comparing results

In the case of multiple-wave solutions, in general, the standard method generates an unbounded term in the first-order correction, hence, it makes sense to compare results of Section 6 with those of the standard analysis only in the strict $O(e^0)$ sense, namely, without a perturbation, where they trivially coincide. The exception, is when both Eqs. (4.16) and (4.24) are obeyed, so that Eq. (4.24) holds. Inspection of Eqs. (4.8), (4.11), (6.5)-(6,9), reveals that the two alternatives coincide, because $a$, $b$ and $m$ obtain the same values in both, and Eq. (4.24) implies $r = s = 0$. That is, the coefficients of the terms that need to be computed numerically in the alternative of Section 6 vanish. The only case, in which a meaningful comparison can be made, is that of a single-wave zero-order approximation. Both the standard analysis and our alternative generate a valid first-order term, $u^{(1)}$, for Eq. (4.2). Consider, therefore, the single-wave case with one vanishing boundary value. The zero-order approximation is given by the $M = 1$ version of Eq. (4.6). The velocity update in the standard analysis is also given by Eq. (4.6), while the alternative of Section 6 is updated according to Eq. (6.10). The standard analysis first-order term, $u^{(1)}$, is given by Eqs. (4.7) and (4.8). In our alternative, it is given by Eqs. (5.1), (6.5)-(6.8) and (6.11). (The latter is solved numerically).

To compare the two solutions, we have to impose boundary values on the solution of Eq. (4.1):

$$w(t,x) \to \begin{cases} 0, & x \to -\infty \\ K > 0, & x \to +\infty \end{cases} \tag{7.1}$$

The limits of the approximate solution for large $|x|$ are:

$$u + e u^{(1)} \to \begin{cases} 0 & x \to -\infty \\ k + e a k^2 & x \to +\infty \end{cases} \tag{7.2}$$

from Eqs. (7.1) and (7.2), one obtains

$$k = K - e a K^2 + O(e^2) \tag{7.3}$$

As a result, in terms of the boundary value, the updated velocity is given by

$$v = k + e\,m k^2 + O(e^2) = K + e(m-a)K^2 + O(e^2) \tag{7.4}$$

Here **m** is the coefficient of $S_3$ in the Normal Form, Eq. (4.3):

$$m = \begin{cases} a_4 & \text{Standard analysis} \\ a_1 + a_2 - a_3 & \text{Section 6} \end{cases} \tag{7.5}$$

The coefficient, $a$, is given by

$$a = \begin{cases} -\frac{1}{2}(2a_1 - a_2 + a_3 - 2a_4) & \text{Standard analysis} \\ \frac{3}{2}a_2 - \frac{3}{2}a_3 & \text{Section 6} \end{cases} \tag{7.6}$$

With Eqs. (7.5) and (7.6), the velocity updates coincide:

$$v = K + e\left(a_1 - \tfrac{1}{2}a_2 + \tfrac{1}{2}a_3\right)K^2 + O(e^2) \tag{7.7}$$

It is easy to show that, the result of Eq. (7.7) for the velocity update is invariant to the (mutually dependent) choices of $a$ and **m**

Having imposed the boundary conditions, we compare the zero-order approximations of the two alternatives. Fig. 4 shows that, as expected, they differ by an $O(e \cdot a \cdot k^2)$ term. Thus, both alternatives generate equally valid zero-order approximates. Fig 5 shows a comparison between the approximate solutions computed through $O(e)$. We see that where the solutions vary slowly, the difference if of $O(e^2 \cdot a^2 \cdot k^4)$, whereas along the characteristic line of the front the error is still $O(e \cdot k^2)$. This is not surprising, as, to guarantee a uniformly valid $O(e^2 \cdot a^2 \cdot k^4)$ error, one must solve the Normal Form, Eq. (4.3), through second order, which we have not done.

## 8. Asymptotic behavior along semi-infinite fronts in multiple-wave case

The choice that converted Eq. (5.2) into Eq. (6.11) has the advantage of rigorously eliminating the obstacle to integrability and the unbounded term in $u^{(1)}$ of Eq. (4.2). However, it requires that a non-negligible contribution to $u^{(1)}$ is computed numerically. In this Section, we offer another alternative, in which the numerically computed corrections become exponentially small as the distance from the collision region of the fronts grows. This new analysis is confined to the behavior of the solution away from the collision region. The property of the zero-order solution that comes to our aid is the observation made in Section 2, that, away from that region, the zero-order solution asymptotes into well-separated single fronts.

Returning to Eqs. (5.1), we note again that, in the multiple-wave case, the term $q \cdot u_x$ is unbounded along the characteristic lines of each of the $(M - 1)$ inelastically generated fronts. To be specific, consider the two-wave case. To avoid the unbounded nature of $q \cdot u_x$, we re-write Eq. (5.1) as

$$u^{(1)}[u;t,x] = au^2 + b(q - q_0)u_x + cu_x + w(t,x) \tag{8.1}$$

In Eq. (8.1), $q_0$ is the value of $q(t,x)$ along the characteristic line of the inelastically generated front (see Eq. (4.21)). The term $(q - q_0) u_x$ is bounded, because the leading linear component in $q$ has been eliminated. By re-writing Eq. (5.1) in the form of Eq. (8.1), one more term is added to Eq. (4.4), which in the notation used in Eq. (5.2) now becomes

$$U_1 + \partial_t w = 2\partial_x(uw) + \partial_x^2 w + \boldsymbol{m} S_3[u] + \boldsymbol{n} R_{21}[u] + b\frac{dq_0}{dt}u_x + \boldsymbol{r}\partial_x(uu_x) + \boldsymbol{s}\,\partial_x S_2[u] \tag{8.2}$$

The last two terms in Eq. (8.2) generate in $u^{(1)}$ finite contributions that persist asymptotically along the inelastically generated fronts (see Section 7) and can be only computed numerically. To eliminate such contributions, we now require $\boldsymbol{r} = \boldsymbol{s} = 0$. Using Eqs. (5.3)-(5.6), the coefficients $\boldsymbol{m}$ and $n$ recover their values in the standard analysis, ($m = a_4$, $n = g = 2a_1 - a_2 - 2a_3 + a_4$) and $a$ and $b$ obtain their values in the *single-wave* case, given in Eq. (4.8). Eq. (8.2) is reduced to

$$U_1 + \partial_t w = 2\partial_x (u w) + \partial_x^2 w + mS_3[u] + nR_{21}[u] + b\frac{dq_0}{dt}u_x \qquad (8.3)$$

Provided the Normal Form is constructed from symmetries only, away from the collision region, the solution of the zero-order term asymptotes into the three well-separated single fronts. Two of them are single fronts with one vanishing boundary value, the parameters of which are not affected by the presence of the other wave. Each evolves along the characteristic line of one of the waves ($x \sim -k_i t - x_{i,0}$). The last two terms in Eq. (8.3) vanish exponentially fast along these two fronts. First, as noted in Section 3, $R_{21}$ vanishes explicitly in the case of a pure single-front solution with one vanishing boundary condition. Hence, it is expected to vanish exponentially along a branch that asymptotes to such a front. As for $q_0$, by its definition, it vanishes exponentially fast along these two fronts. As a result, along these two fronts, the Normal Form, Eq. (4.3), becomes

$$u_t = S_2[u] + e\, a_4\, S_3[u] + O(e^2) \qquad (8.4)$$

and Eq. (8.3) is reduced to

$$\partial_t w = 2\partial_x (u w) + \partial_x^2 w + e.s.t \qquad (8.5)$$

Thus, away from the collision region, in triangular wedges in the *x-t* plane, these two fronts are analyzed just as in the single-front case, with exponentially small numerical corrections.

The inelastically generated semi-infinite front asymptotes to a single-wave front with both boundary values non-vanishing, the structure of which is given in Eq. (2.3), with $u_m = k_1$ and $u_p = k_2$. The effect of the perturbation on such a front is given by Eq. (4.10). The only change in the zero-order solution is that its velocity is updated from the unperturbed value into

$$v = k_1 + k_2 + e\{(k_1 k_2)(a_1 + a_2 - a_3) + (k_1^2 + k_2^2)a_4\} \tag{8.6}$$

In the following, we show that Eq. (8.6) is, indeed, obtained. Both last terms in Eq. (8.3) become proportional to the symmetry $S_1 = u_x$ along the characteristic line of the inelastically generated front of Eq. (2.12). For $R_{21}$, this is shown in Eq. (6.2). In the last term, $dq_0/dt$ asymptotes to a constant, $-k_1 \cdot k_2$, hence,

$$\frac{dq_0}{dt} u_x \xrightarrow[k_1 k_2 t \to \infty]{} -k_1 k_2 u_x + e.s.t. \tag{8.7}$$

As a result, Eq. (8.3) becomes

$$U_1 + \partial_t w = 2\partial_x(u w) + \partial_x^2 w + a_4 S_3 + k_1 k_2 (a_1 + a_2 - a_3 - a_4) u_x + e.s.t. \tag{8.8}$$

To avoid secular terms, one shifts the symmetry part to the Normal Form, which, becomes

$$u_t = S_2[u] + e(a_4 S_3 + k_1 k_2 (a_1 + a_2 - a_3 - a_4) u_x) + O(e^2) \tag{8.9}$$

leading to the solution of Eq. (2.3) with $u_m = k_1$ and $u_p = k_2$, and a velocity updated as in Eq. (8.6).

Eq. (8.8) is reduced to one that generates an exponentially small numerical correction to $u^{(1)}$:

$$\partial_t w = 2\partial_x(u w) + \partial_x^2 w + e.s.t. \tag{8.10}$$

The extension to solutions with $M = 2$ waves is straightforward. The two elastically generated fronts are analyzed in the same manner as they are analyzed in the two-wave case. The $(M-1)$ inelastically generated fronts are analyzed in the same manner as in the example discussed above, with $k_1$ and $k_2$ replaced by $k_i$ and $k_{i+1}$, $1 < i < M - 1$ for each front.

## 9. Concluding comments

In this paper, we have shown how the freedom in the perturbative expansion can be exploited in the case of the perturbed Burgers equation to shift the loss of asymptotic integrability from the Normal Form to the higher order terms in the Near Identity Transformation, and to avoid the occurrence of unbounded terms in higher-orders in the expansion. The price paid is that part of the contribution to the expansion of the solution has to be computed numerically. The gain is that the Normal Form remains asymptotically integrable, so that the zero-order solution retains the simple nature of the solution of the unperturbed equation. Identifying the role played by inelastically generated fronts in the emergence of both unbounded terms and of obstacles to integrability was the key to overcoming both deficiencies of the standard analysis.

Obstacles to integrability are also encountered in the standard analysis of other perturbed evolution equations, see, e.g., [7,8,11,12], where the perturbed KdV and NLS have been studied. The unperturbed equation of these systems has soliton solutions. The latter are localized, unlike the fronts of the Burgers equation. As a result, although the formalism allows terms of the type of $qu_x$, they are bounded. Moreover, the solutions of the KdV and NLS satisfy conservation laws that do not have counterparts in the case of the Burgers equation. As a result, the solitons in multi-soliton solutions of the unperturbed equations, and of the Normal Form interact inelastically only in a small domain around the origin in the $x$-$t$ plane (the interaction region), where the solitons lose their individual identity. Away from that region, the solitons regain their original structure (except for a possible phase shift). As inelastic effects do not extend beyond the bounded interaction region, the freedom in the Normal Form expansion can be exploited so as to confine the effect of the obstacles to integrability to the interaction region, guaranteeing that no superfluous unbounded terms emerge. WE have studied the case of the perturbed KdV equation. Some of the results are published in [19, 20].

FIGURE CAPTIONS

Fig.1 Solutions of Eq. (2.1) a) $u(t,x)$ – single wave $k = 0.5$, $x_{1,0} = 0$; b) $?_x u$ - single-wave; b) $u(t,x)$ – three waves; $k_1 = 5$, $k_2 = -5$, $k_3 = -2$, $x_{1,0} = 1$, $x_{2,0} = 2$, $x_{3,0} = 3$; d) $?_x u$ – three waves. Note that because $x_{i,0} ? 0$, the fronts collide at diferent points.

Fig.2 Solution of Eq. (6.11), single-wave case, $k = 5$, with driving term: a) $?_x(u \cdot u_x)$; b) $?_x S_2[u]$.

Fig. 3 Solution of Eq. (6.11), two-wave case, $k = 5$, with driving term: a) $?_x(u \cdot u_x)$; b) $?_x S_2[u]$.

Fig. 4 Difference between zero-order approximation of standard analysis (Section4) and alternative of Section 6 for single-front case. $e = 0.001$, $K = 5$ (see Eq. (7.1)), $\boldsymbol{a}_1 = 1$, $\boldsymbol{a}_2 = \boldsymbol{a}_3 = 2$, $\boldsymbol{a}_4 = 3$.

Fig. 5 Difference between first-order approximation of standard analysis (Section4) and alternative of Section 6 for single-front case. Paraameters as in Fig. 4.

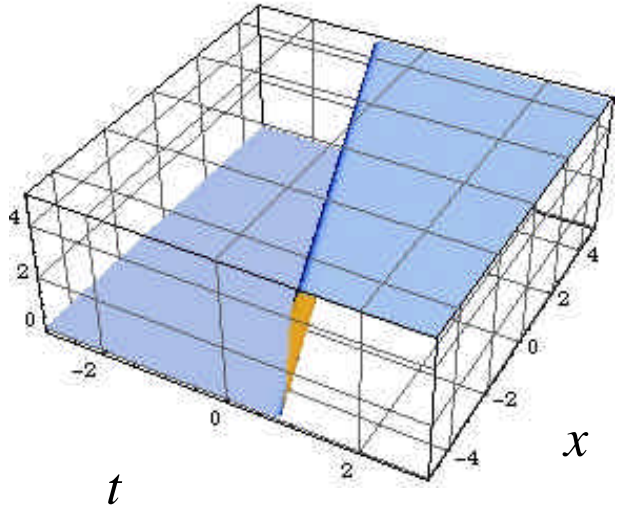

Fig. 1a

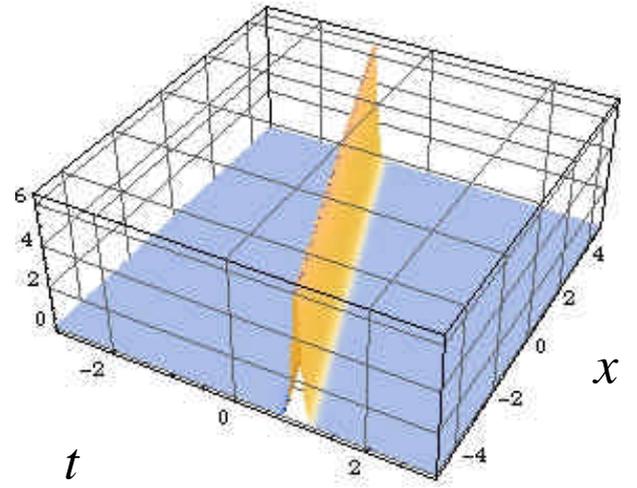

Fig. 1b

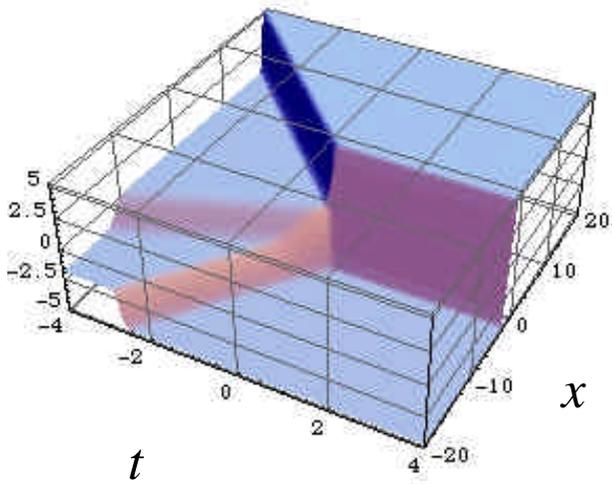

Fig. 1c

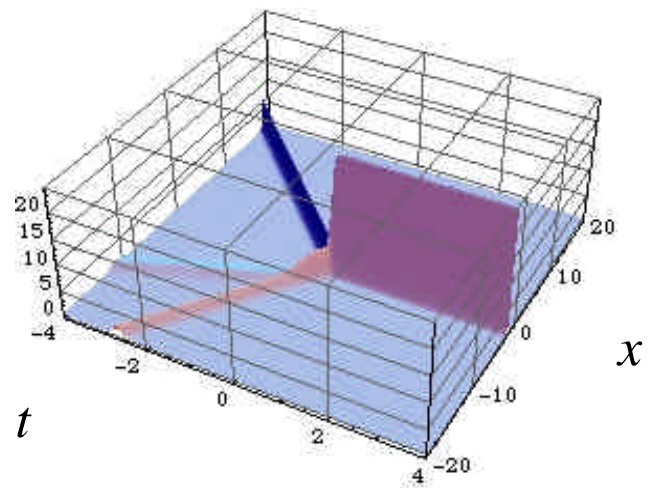

Fig. 1d

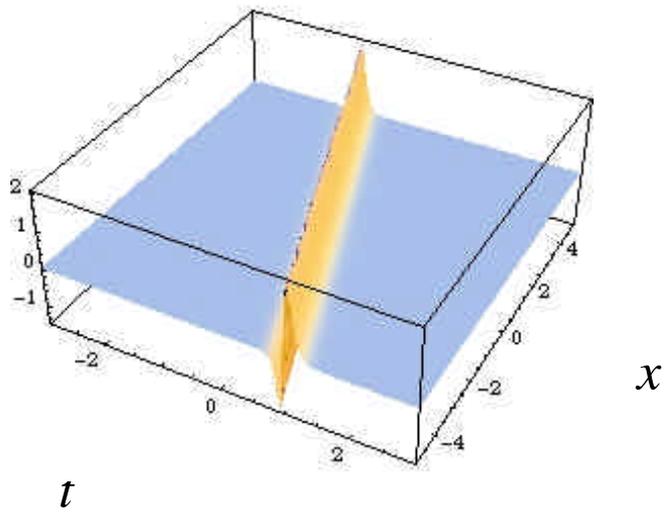

Fig. 2a

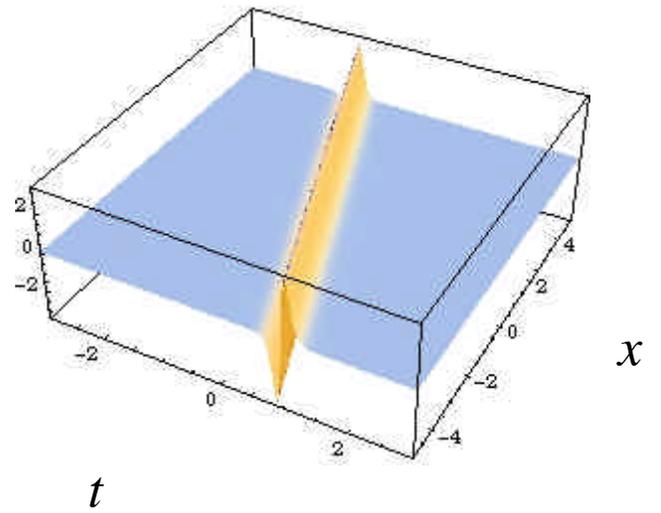

Fig. 2b

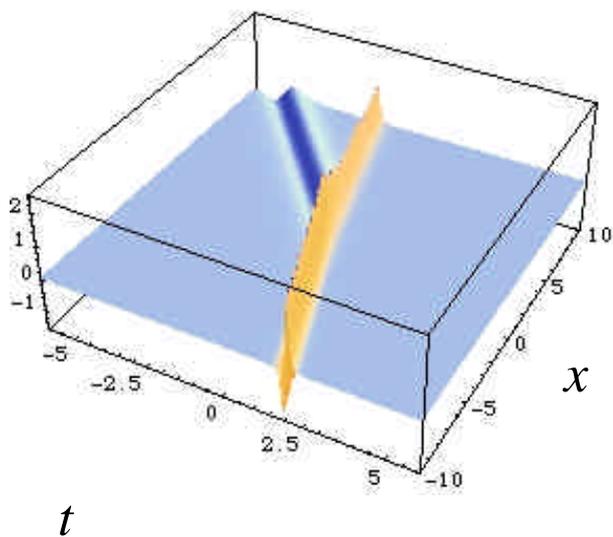

Fig. 3a

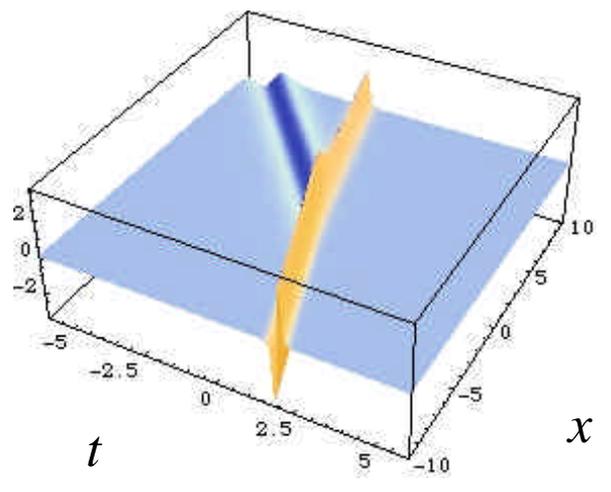

Fig. 3b

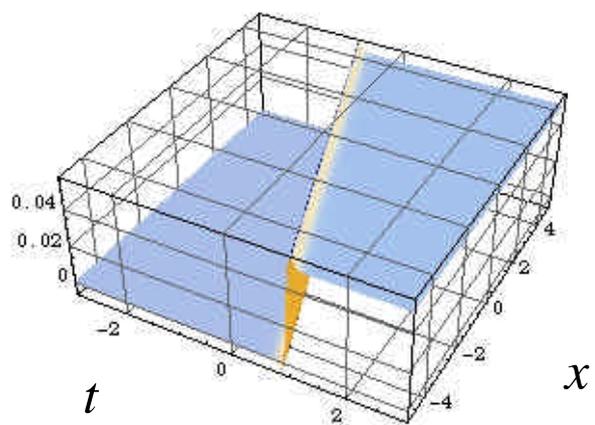 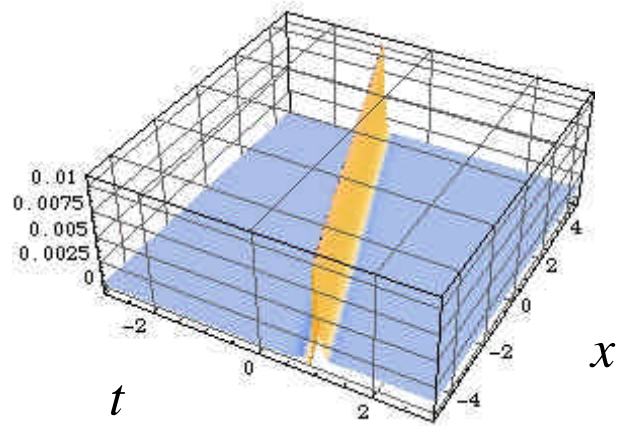

Fig. 4 Fig. 5